 \documentclass[final,3p,times,sort&compress]{elsarticle}

\usepackage{multirow,setspace,times,amssymb,amsmath,graphicx,color,rotating,subfigure,url}
\usepackage{lineno}
\usepackage{natbib}

\journal{Physica A}

\begin{document}

\begin{frontmatter}



\title{Analyzing the prices of the most expensive sheet iron all over the world:\\Modeling, prediction and regime change}
\author[BS,RCE]{Fu-Tie Song}
\author[BS,RCE,SS,RCSE,RCFE]{Wei-Xing Zhou \corref{cor}}
\cortext[cor]{Corresponding author. Address: 130 Meilong Road, P.O.
Box 114, School of Business, East China University of Science and
Technology, Shanghai 200237, China, Phone: +86 21 64253634, Fax: +86
21 64253152.}
\ead{wxzhou@ecust.edu.cn} %

\address[BS]{School of Business, East China University of Science and Technology, Shanghai 200237, China}
\address[RCE]{Research Center for Econophysics, East China University of Science and Technology, Shanghai 200237, China}
\address[SS]{School of Science, East China University of Science and Technology, Shanghai 200237, China}
\address[RCSE]{Engineering Research Center of Process Systems Engineering (Ministry of Education), East China University of Science and Technology, Shanghai 200237, China}
\address[RCFE]{Research Center on Fictitious Economics \& Data Science, Chinese Academy of Sciences, Beijing 100080, China}

\begin{abstract}
The private car license plates issued in Shanghai are bestowed the title of ``the most expensive sheet iron all over the world'', more expensive than gold. A citizen has to bid in an monthly auction to obtain a license plate for his new private car. We perform statistical analysis to investigate the influence of the minimal price $P_{\min}$ of the bidding winners, the quota $N_{\rm{quota}}$ of private car license plates, the number $N_{\rm{bidder}}$ of bidders, as well as two external shocks including the legality debate of the auction in 2004 and the auction regime reform in January 2008 on the average price $P_{\rm{mean}}$ of all bidding winners. It is found that the legality debate of the auction had marginal transient impact on the average price in a short time period. In contrast, the change of the auction rules has significant permanent influence on the average price, which reduces the price by about 3020 yuan Renminbi. It means that the average price exhibits nonlinear behaviors with a regime change. The evolution of the average price is independent of the number $N_{\rm{bidder}}$ of bidders in both regimes. In the early regime before January 2008, the average price $P_{\rm{mean}}$ was influenced only by the minimal price $P_{\min}$ in the preceding month with a positive correlation. In the current regime since January 2008, the average price is positively correlated with the minimal price and the quota in the preceding month and negatively correlated with the quota in the same month. We test the predictive power of the two models using 2-year and 3-year moving windows and find that the latter outperforms the former. It seems that the auction market becomes more efficient after the auction reform since the prediction error increases.
\end{abstract}


\end{frontmatter}


\section{Introduction}
\label{S1:Introduction}

Gold, silver, platinum and palladium are precious metals, in which gold is also called hard currency. Precious metals are certainly expensive. For instance, the price of gold (London Gold Fixing PM) was 810.50 USD per troy ounce on 21 December 2007 and 1106.50 USD per troy ounce on 18 December 2009. The foreign exchange rates between US dollar and Chinese Renminbi (RMB) were 7.3572 RMB/USD and 6.8284 RMB/USD on these two days. Equivalently, the two prices are 194,933 RMB/kg and 246,997 RMB/kg. There is no doubt that gold or platinum cannot be termed as the most expensive metal in the world since there exist metals that are more precious. For most people, it is hard to tell what is the most previous metal in the world. Nevertheless, a large population of Chinese people know about the existence of ``the most expensive sheet iron all over the world'' in Shanghai, which is the nickname given to the private car license plate (PCLP) issued by the Vehicle Administrative Organ of the Traffic Administering Department of the Shanghai Municipal Public Security Bureau. The average auction price of Shanghai's PCLPs was 56,042 RMB on 22 December 2007 as the historical high since January 2000 and 37,593 RMB on 19 December 2009. Since the weight of a license plate is about 0.19 kg, we find that the price of this ``special'' metal was 294,958 RMB/kg on 22 December 2007 and 197,858 RMB/kg on 19 December 2009.


In Shanghai, a private car owner has to bid for a license plate in the monthly auction held by Shanghai International Commodity Auction Co. Ltd. The PCLP auction started since 1986 in Shanghai in order to control the overly fast growth of private cars and relieve the traffic congestion. At the time, the minimum bid price was posed to be 100,000 RMB and the actual price could be as high as 300,000 RMB. In early 1998, a special kind of license plates was released and the minimum bid price reduced to 20,000 RMB, which can only be used for Santana 2000 cars made by Shanghai Volkswagen and later Buicks produced by Shanghai GM. In contrast, the minimum bid price for not-made-in-Shanghai cars remained unchanged. This caused a trade war between Shanghai and Hubei province. This regional protectionism policy was abolished in January 2000 and the minimum bid price of auction was canceled as well, which applied to domestic cars while a minimum bid price of about 50,000 RMB still applied to imported cars. The minimum bid price for imported cars was canceled in March 2003, one year and a half after China became a member of the World Trade Organization. Before 2008, the auction was blind and the bidders were allowed to submit a price only once. Since January 2008, the policy changed with an open auction system taking into effect and the bidders can revise their submitted bid price twice.

The legality of the private car license plate auction was questioned from time to time \cite{Editorial-2004-cnEM}. The debate attracted most attention of the public happened in 2004 and culminated in May \cite{Hu-2004-cnCSP}. In April 2004, the Beijing Youth Daily reported that the National Development and Reform Commission would stop Shanghai's auction and the Focal Talk Online of Xinhua Net questioned the high price of a license plate. On May 24, Mr. Hai Huang, the Assistant Minister of the Ministry of Commerce of China, explicitly expressed that the PCLP auction violated the Law of the People's Republic of China on Road Traffic Safety that came into force on 1 May 2004. On the second day, the spokeswoman of Shanghai Municipal Government replied that Shanghai would not change the practice of PCLP auction for the moment. The majority of the domestic legal community argued that Shanghai's PCLP auction is illegal \cite{Zheng-Lu-2005-cnPAL,Yue-Fan-2005-cnJNCPI,Yang-Zhang-2008-cnJPSCPCCCMC}, while some others disagreed \cite{Yang-Huang-2005-cnALR}.

The function of the auction is more than what the officials claimed. From October 1998 to October 2001, there was a policy of combined house-car sale with an extra bonus of a free PCLP. This is certainly not the only reason that fueled the real estate bubble in Shanghai. However, it highlights the gestation of the on-going housing bubble \cite{Zhou-Sornette-2004a-PA}. A direct consequence of the PCLP auction is that a large proportion of the private car owners equip their cars with license plate issued by other provinces, which not only denies the initial intent of the auction but also causes other administrative problems. There is no clear scientific evidence showing whether the auction does improve the traffic situation in Shanghai. More generally, there are only a few quantitative studies on the topic of PCLP auction. Lou and Wang studied the vehicle license auction based on the private value bid model and found that the reform of the auction in January would lower the minimal price of the winning bidders in a short period \cite{Luo-Wang-2009-cnSTE}. Alternatively, Xu tried to predict the average price based on a simple linear regression model with a dummy variable \cite{Xu-2005-ET}. Most studies focus on the physical properties of complex road traffic based on numerical modeling and real data \cite{Tang-Huang-Gao-2005-PRE,Wu-Gao-Sun-Huang-2006-EPL,Tang-Huang-Gao-Wong-2007-PA,Tang-Huang-Wong-Gao-Zhang-2009-CTP,Xie-Gao-Zhao-Li-2009-PA,Yang-Gao-Zhao-Si-2009-TRR}, which provide insights into the understanding of traffic congestion and possible solutions to optimize traffic design.

In this work, we attempt to investigate the behavior of the monthly average price $P_{\rm{mean}}$ of the winning bidders of private car license plates in Shanghai. In Section \ref{S1:Data}, we describe the data sets and their basic statistical properties. Section \ref{S1:Model} performs systematical analysis of different regressive models to understand what are the influencing factors of the average price. In Section \ref{S1:Prediction}, we try to construct and select predictive models using moving windows, which enables us to identify a change of regime occurred in January 2008. Section \ref{S1:Summay} summarizes.

\section{Data sets}
\label{S1:Data}

The data sets we analyze in this work are retrieved from the web site of Shanghai International Commodity Auction Co. Ltd\footnote{http://www.alltobid.com/guopai/, accessed on 6 January 2010.}. There are four variables, the average price $P_{\rm{mean}}$ of all bidding winners, the minimal price $P_{\min}$ of the bidding winners, the quota $N_{\rm{quota}}$ of private car license plates, and the number $N_{\rm{bidder}}$ of bidders. The data are monthly and cover the time period from January 2002 to December 2009. No auction occurred in February 2008. Therefore, the length of each data set is 95. Table \ref{TB:STAT} depicts the basic statistics of the four variables.

\begin{table}[htb]
 \centering
 \caption{\label{TB:STAT} Basic statistics of the four variables.}
 \medskip
 \begin{tabular}{ccccccccccccc}
 \hline\hline
    Variable        & Minimum &  Median & Maximum &   Mean & Std. Dev. & Skewness & Kurtosis  \\\hline
  $P_{\rm{mean}}  $ &   14057 &   34842 &   56042 &   33950 &    8389 &  -0.099 &   3.249\\
  $P_{\min}$        &     100 &   33800 &   53800 &   31938 &    9626 &  -0.555 &   3.692\\
  $N_{\rm{quota}} $ &    1400 &    5690 &   16000 &    5628 &    2060 &   1.140 &   8.257\\
  $N_{\rm{bidder}}$ &    3525 &   10170 &  110234 &   12846 &   12657 &   5.611 &  40.662\\
  \hline\hline
  \end{tabular}
\end{table}

\begin{figure}[htb]
 \centering
 \includegraphics[width=8cm]{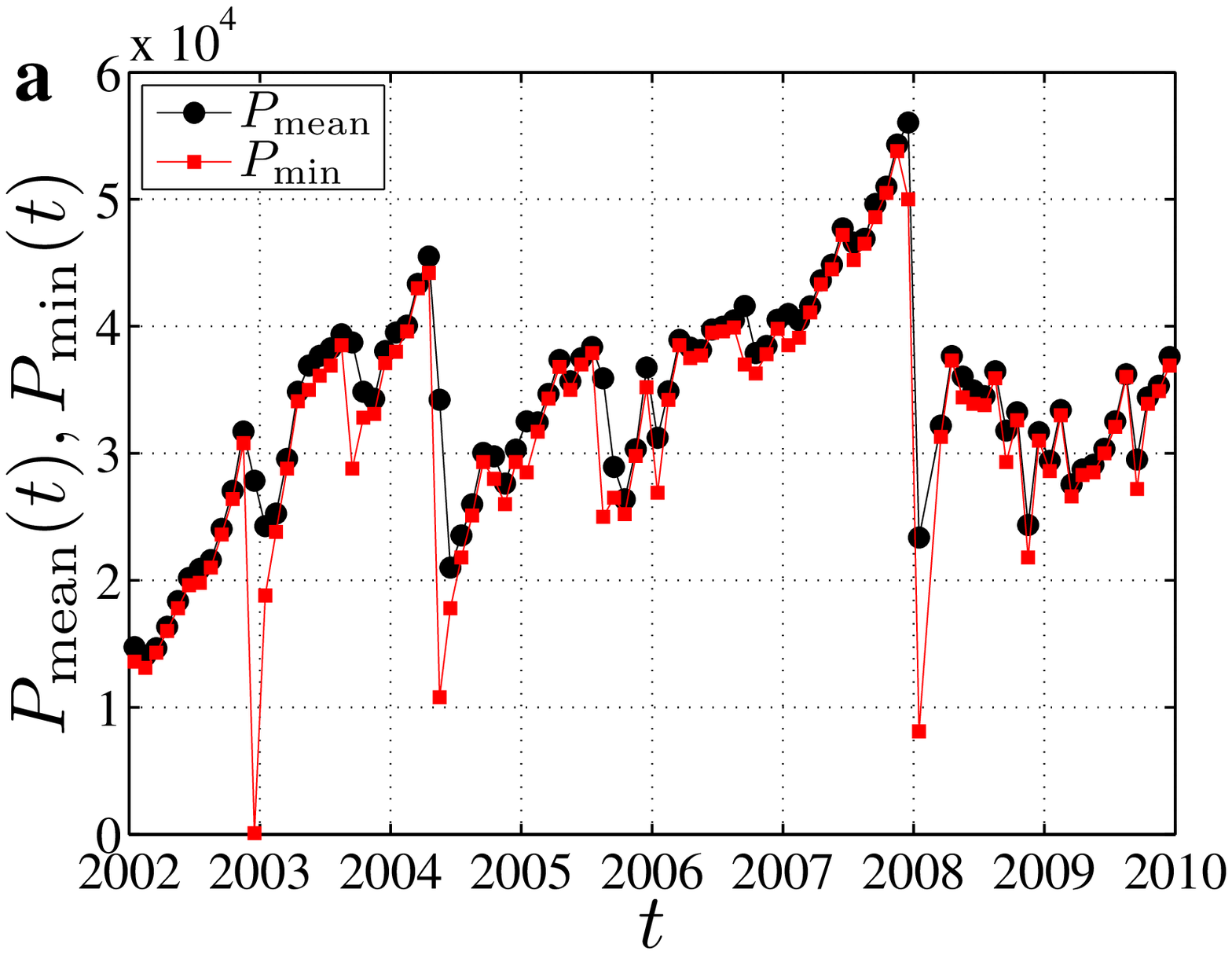}
 \includegraphics[width=8cm]{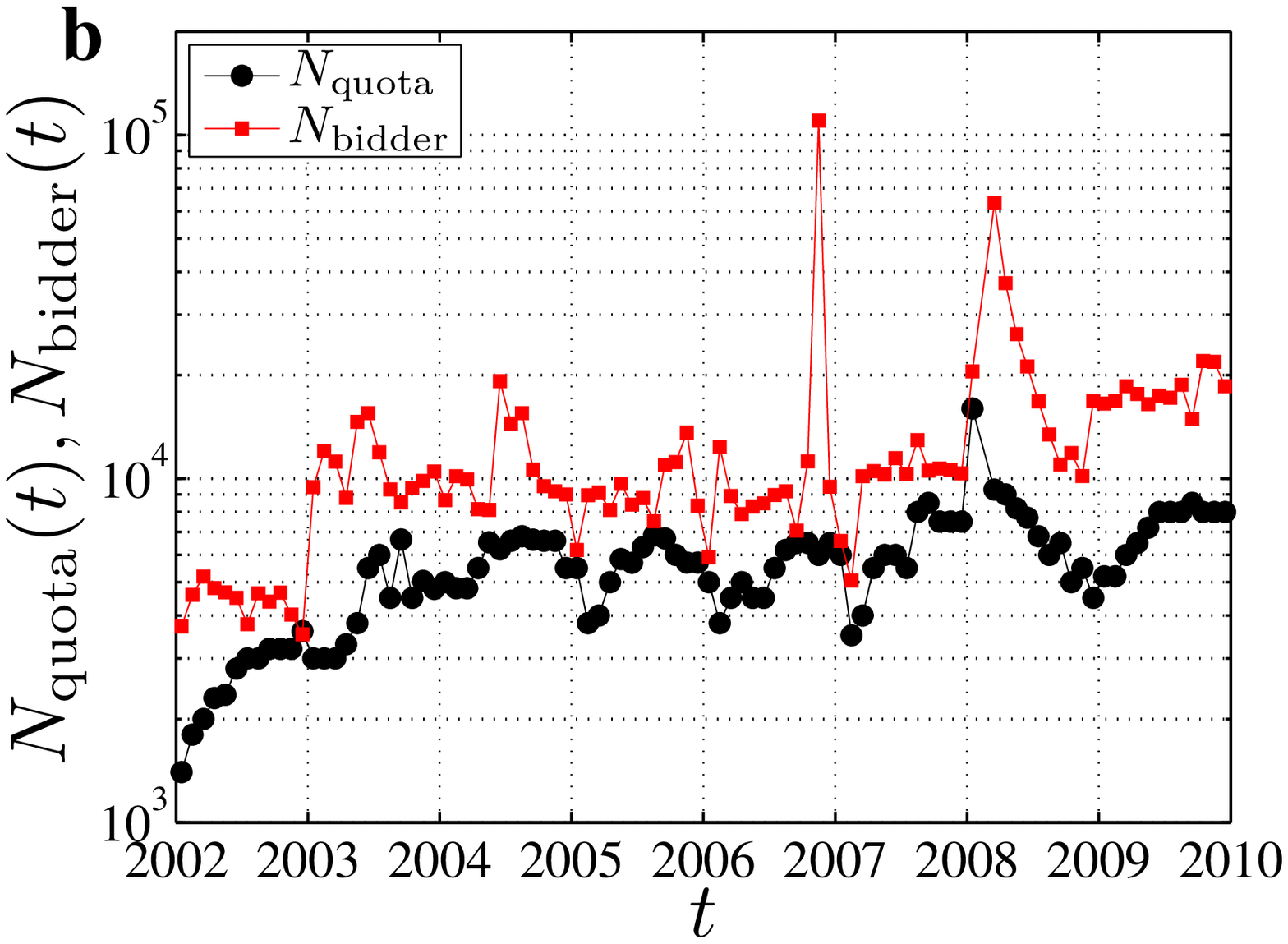}
 \caption{\label{Fig:Price:Number} (Color online) Evolution of the average price $P_{\rm{mean}}$ of all bidding winners, the minimal price $P_{\min}$ of the bidding winners, the quota $N_{\rm{quota}}$ of private car license plates, and the number $N_{\rm{bidder}}$ of bidders.}
\end{figure}

The evolution of the four variables is illustrated in Fig.~\ref{Fig:Price:Number}. According to Fig.~\ref{Fig:Price:Number}a, we observe three time periods separated by two months, May 2004 and January 2008. In each of the first two periods, the average price has an increasing trend. Since January 2008, there is no evident trend and the price becomes more stable. The minimal price curve is on average close to the average price curve. However, we see three striking downward spikes occurred in December 2002, May 2004 and January 2008. For the number of bidders, there is also a weak increasing trend with a marked spike in October 2006 and a peak around March 2008. The monthly quota also increases with a strong yearly periodic pattern. In each year except 2008, the quotas in the early months are low, which is caused by the fact that the sale of private cars is low in these months. It is not clear how the Shanghai Municipal Government determines the quota in each month.

\begin{figure}[htb]
 \centering
 \includegraphics[width=7.9cm]{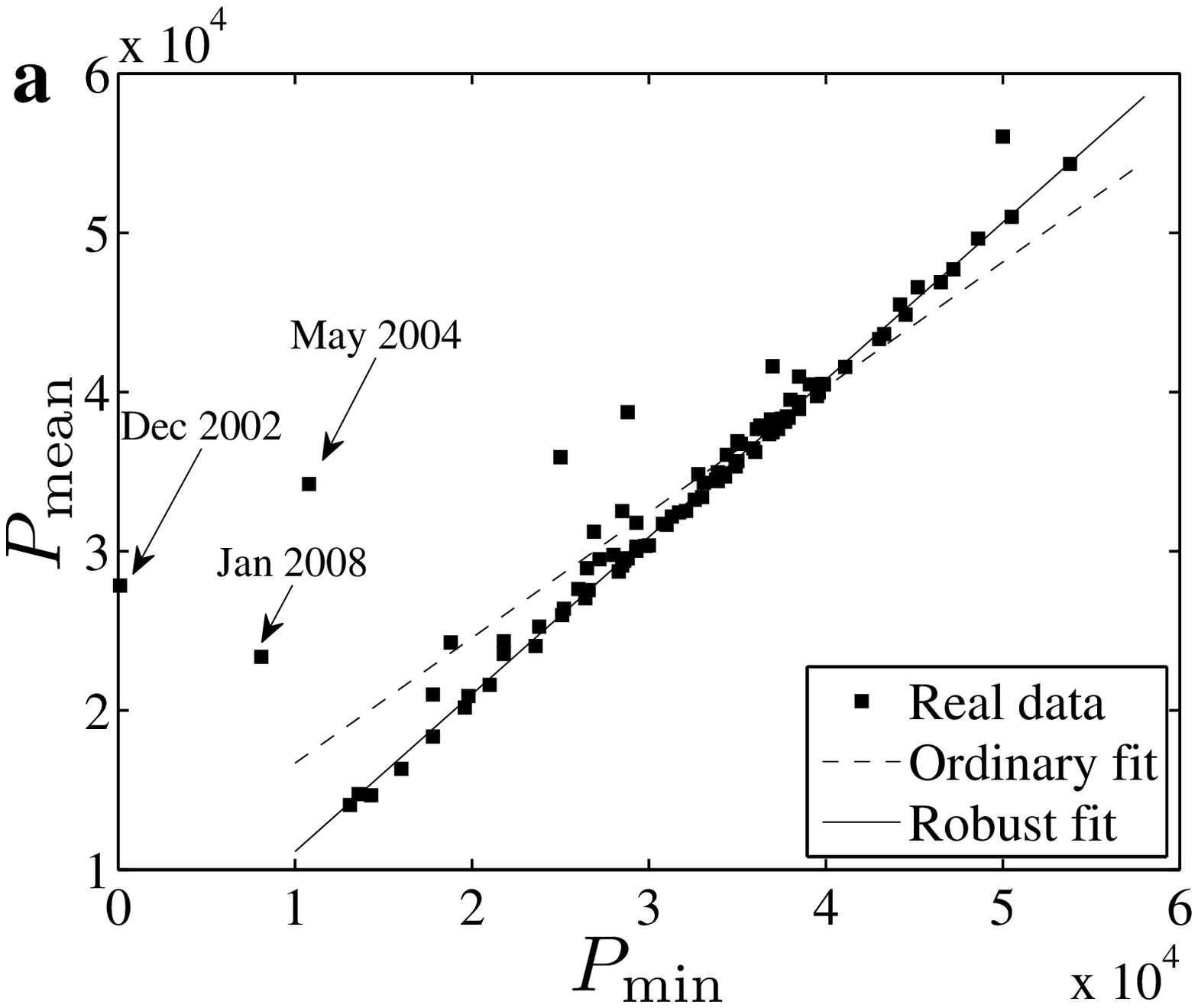}
 \includegraphics[width=8.0cm]{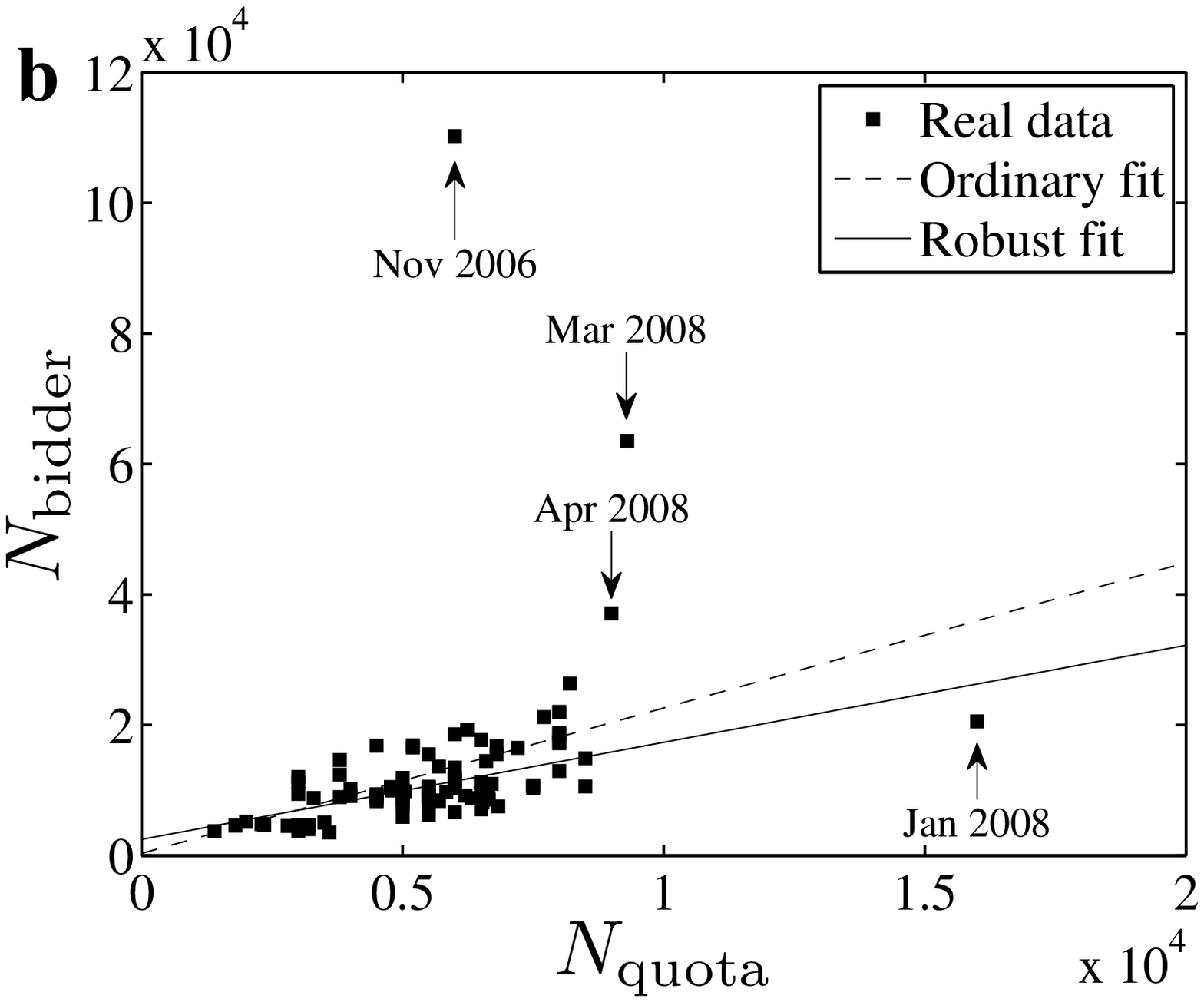}
 \caption{\label{Fig:Corr} (a) Positive correlation between the average price $P_{\rm{mean}}$ of all bidding winners and the minimal price $P_{\min}$ of the bidding winners. (b) Positive correlation between the quota $N_{\rm{quota}}$ of private car license plates and the number $N_{\rm{bidder}}$ of bidders.}
\end{figure}

Fig.~\ref{Fig:Corr}a shows the scatter plot of the average price $P_{\rm{mean}}$ against the minimal price $P_{\min}$, in which all the data points locate above the diagonal line $P_{\rm{mean}}=P_{\min}$. A nice linear relationship is observed:
\begin{equation}
 P_{\rm{mean}}(t) = a_0 + a_1 P_{\min}(t).
\end{equation}
An ordinary linear regression gives that $a_0=8813 \pm 2570$ and $a_1=0.787\pm0.077$, where the errors are determined at the 95\% confidence level. The R-square is 0.816. According to Ref.~\cite{Chatterjee-Hadi-1986-SS}, six outliers are identified at the 5\% significance level including December 2002, September 2003, May 2004, August 2005, January 2008, and March 2008, three of which are indicated in Fig.~\ref{Fig:Corr}a. We further perform a robust fit using iteratively re-weighted least squares \cite{Holland-Welsch-1977-CStm,Huber-1981,Street-Carroll-Ruppert-1988-AS}, and find that $a_0=1251 \pm 332$ and $a_1=0.988\pm0.010$ at the 95\% confidence level. The two fits are shown in Fig.~\ref{Fig:Corr}a for comparison. In Fig.~\ref{Fig:Corr}b, there is also a linear relationship between the two variables:
\begin{equation}
 N_{\rm{bidder}}(t) = b_0 + b_1 N_{\rm{quota}}(t).
\end{equation}
The ordinary linear regression gives that $b_0=298\pm7060$ and $b_1= 2.229\pm1.179$ with the R-square being 0.132. At the 5\% significance level, two outliers are identified in November 2006 and March 2008. In contrast, the robust regression gives that $b_0=2463\pm1325$ and $b_1=1.488\pm0.221$. The positive correlation between $N_{\rm{bidder}}$ and $N_{\rm{quota}}$ means that the number of bidders is strongly influenced by the quota in the same month, which is publicly announced before the auction.

\section{Modeling the evolution of average price}
\label{S1:Model}

\subsection{Basic statistical analysis}

In our analysis, we attempt to model the average price $P_{\rm{mean}}(t)$. The first explanatory variable is the minimal price $P_{\min}(t-1)$ in the preceding month. Rationally, the supply $N_{\rm{quota}}$ and demand $N_{\rm{bidder}}$ of the PCLPs are expected to have essential influence in the formation of the price. Before the auction in month $t$, the supplied quota $N_{\rm{quota}}(t)$ of license plates is released by the Shanghai Municipal Government. However, the number of bidders $N_{\rm{bidder}}(t)$ is known only after the auction finishes. Therefore, we add $N_{\rm{quota}}(t)$ and $N_{\rm{bidder}}(t-1)$ as two explanatory variable. Since the demand $N_{\rm{bidder}}(t-1)$ in month $t-1$ is included, we also add the supply $N_{\rm{quota}}(t-1)$ in month $t-1$ in the model. We thus obtain the following model
\begin{equation}
 P_{\rm{mean}}(t) = c_0+c_1P_{\min}(t-1)+c_2N_{\rm{quota}}(t)+c_3N_{\rm{quota}}(t-1)+c_4N_{\rm{bidder}}(t-1)+\epsilon(t),
 \label{Eq:Model:1}
\end{equation}
where $\epsilon(t)$ is the residual.

We calibrate the model (\ref{Eq:Model:1}) to all the data from January 2002 to December 2009. The estimates of the coefficients are $c_0= 8601$, $c_1=0.814$, $c_2=-2.343$, $c_3=2.243$, and $c_4=0.025$, respectively. The R-square is 0.81. Statistical test shows that $c_0$, $c_1$, $c_2$ and $c_3$ are significantly different from zero with all the $p$-values equal to zero. In contrast, the $p$-value of $c_4$ is 0.44, showing that the influence of the bidder number in the previous month $N_{\rm{bidder}}(t-1)$ on the average price $P_{\rm{mean}}(t)$ is insignificant. Discarding the variable $N_{\rm{bidder}}$ from model (\ref{Eq:Model:1}), we have
\begin{equation}
 P_{\rm{mean}}(t) = c_0+c_1P_{\min}(t-1)+c_2N_{\rm{quota}}(t)+c_3N_{\rm{quota}}(t-1)+\epsilon(t).
 \label{Eq:Model:2}
\end{equation}
A linear least-squares regression gives that $c_0= 8584$, $c_1=0.814$, $c_2=-2.316$, and $c_4=2.277$. The R-square is 0.809. Statistical test shows that all the coefficients are significantly different from zero.

Table \ref{TB:Model1:Model2} summarizes the results of model calibration. One can see that the corresponding estimates for the same variable in the two models are very close to one another. In addition, introducing $N_{\rm{bidder}}$ in model (\ref{Eq:Model:1}) gives marginal improvement in the explanatory power of the model since the value of $R^2$ increases only slightly from 0.809 to 0.810.

\begin{table}[htb]
\centering
\caption{Calibration of Model (\ref{Eq:Model:1}) and Model (\ref{Eq:Model:2}).  An estimate followed by ``***'' means that the variable is significant at the 0.001 level. An estimate followed by ``**'' means that the variable is significant at the 0.01 level. An estimate followed by ``*'' means that the variable is significant at the 0.1 level.}
\label{TB:Model1:Model2}
\medskip
\begin{tabular}{ccccccccccccccccccccccccc}
  \hline\hline
   Model & $R^2$  & $c_0$ & $c_1$ & $c_2$ &  $c_3$ & $c_4$ \\\hline
  (\ref{Eq:Model:1})  & 0.810 &  8601*** & 0.814*** & -2.343*** & 2.243*** & 0.025 \\
  (\ref{Eq:Model:2})  & 0.809 &  8584*** & 0.814*** & -2.316*** & 2.277*** &  /    \\
  \hline\hline
\end{tabular}
\end{table}

\subsection{Influence of exogenous shocks on the average price}

We now take into account two exogenous shocks that might have influence on the average price. The first shock is the debate on the legality of the auction in 2004. The ministry of Commerce of the central government argued that the PCLP auction is illegal, which was supported by many law scholars. The controversy peaked in May 2004 and many citizens in Shanghai supposed that the auction would be canceled soon. This expectation made the price drop dramatically in May 2004, as shown in Fig.~\ref{Fig:Price:Number}. The second exogenous shock is the change of the auction rules that took into effect in January 2008, which also remarkably pulled down the price in January 2008. We thus introduce two dummy variables, $D_{2004}$ and $D_{2008}$. When the time is earlier than May 2004, we set $D_{2004}=0$; otherwise, $D_{2004}=1$. When the time is earlier than January 2008, we set $D_{2008}=0$; otherwise, $D_{2008}=1$. The model (\ref{Eq:Model:1}) becomes
\begin{equation}
 P_{\rm{mean}}(t) = c_0+c_1P_{\min}(t-1)+c_2N_{\rm{quota}}(t)+c_3N_{\rm{quota}}(t-1)+c_4N_{\rm{bidder}}(t-1)+d_1 D_{2004}+d_2 D_{2008}+\epsilon(t),
 \label{Eq:Model:3}
\end{equation}
The dummy variable is used to explain the possible influence of the exogenous shocks on the average price. If the coefficient $d_1$ or $d_2$ is statistically significant, we can conclude that the behavior of price formation changed owning to the corresponding exogenous shock.

We calibrate the model (\ref{Eq:Model:3}) to all the data from January 2002 to December 2009. The estimates of the coefficients are $c_0= 7925$, $c_1=0.780$, $c_2=-2.009$, $c_3=2.423$, $c_4=0.040$, $d_1= -771$, and $d_2=-3144$. The R-square is 0.830. Statistical test shows that $c_0$, $c_1$, $c_2$ and $c_3$ are significantly different from zero at the 0.1\% level, while the $p$-values are 0.206 for $c_4$, 0.462 for $d_1$, and 0.004 for $d_2$. Again, the number of bidders in the preceding month has no impact on the average price in the current month. In addition, the legality debate does not have permanent impact on the price afterwards. It is reasonable since the Shanghai Municipal Government insisted that it was too early to abolish the auction system, which was still an effective way to alleviate the heavy traffic.

We remove the two insignificant variables $N_{\rm{bidder}}$ and $D_{2004}$ to construct a new model:
\begin{equation}
 P_{\rm{mean}}(t) = c_0+c_1P_{\min}(t-1)+c_2N_{\rm{quota}}(t)+c_3N_{\rm{quota}}(t-1) +d_2 D_{2008}+\epsilon(t).
 \label{Eq:Model:4}
\end{equation}
We find that $c_0= 8144$, $c_1=0.774$, $c_2=-2.007$, $c_3=2.401$, and $d_2=-3020$. The R-square is 0.826. The $p$-value of $d_2$ is 0.005. It is found that $c_0$, $c_1$, $c_2$ and $c_3$ are statistically significant at the 0.1\% level, while $d_2$ is significant at the 1\% level. Table \ref{TB:Model3:Model4} summarizes the results of model calibration. The change of auction rules has permanent impact on the average price, which reduces the average price by 3020 RMB. The obvious reason is that the new auction procedure is more transparent and the bidders can monitor the real-time evolution of the price and revise accordingly their bid prices twice. In other word, the market is more efficient since January 2008.

\begin{table}[htb]
\centering
\caption{Calibration of Model (\ref{Eq:Model:3}) and Model (\ref{Eq:Model:4}). An estimate followed by ``***'' means that the variable is significant at the 0.001 level. An estimate followed by ``**'' means that the variable is significant at the 0.01 level. An estimate followed by ``*'' means that the variable is significant at the 0.1 level.}
\label{TB:Model3:Model4}
\medskip
\begin{tabular}{ccccccccccccccccccccccccc}
  \hline\hline
   Model & $R^2$    & $c_0$ & $c_1$ & $c_2$ &  $c_3$ & $c_4$ &  $d_1$ &   $d_2$ \\\hline
  (\ref{Eq:Model:3})  & 0.830 &  7925*** & 0.780*** & -2.009*** & 2.423*** & 0.040 &  -771 & -3144** \\
  (\ref{Eq:Model:4})  & 0.826 &  8144*** & 0.774*** & -2.007*** & 2.401*** &   /   &   /   & -3020** \\
  \hline\hline
\end{tabular}
\end{table}

\section{Prediction of the average price}
\label{S1:Prediction}

\subsection{Models for prediction}
\label{S2:Model4Predict}

The empirical analysis presented in Section \ref{S1:Model} enables us to understand the dynamics of the average price. In order to make predictions, we need to further study the performance of the proposed models. Rather than using all the data available, it is rational to using the data in a fixed moving window of size $S$ right before the month to be predicted. In this vein, we calibrate a model using data from month $t-S$ to month $t-1$ to obtain the estimates of model parameters, and then extrapolate to predict the average price of month $t$. For this purpose, we have to investigate the ``local'' descriptive power of the models, which might be different from the ``global'' results obtained in  Section \ref{S1:Model}.

We consider two window sizes, $S=24$ months and $S=36$ months. This choice is more or less subjective but not irrational. On the one hand, the window should not be too long, since earlier information is expected to have less influence on the current average price. On the other hand, the window should not be too short, since the models will otherwise be unstable. For the sake of completeness, we investigate model (\ref{Eq:Model:3}), which contains all the variables we considered. In this way, we are able to determine which influencing factors are significant and how the collection of influencing factors changed along time.

\begin{figure}[htb]
 \centering
 \includegraphics[width=8cm]{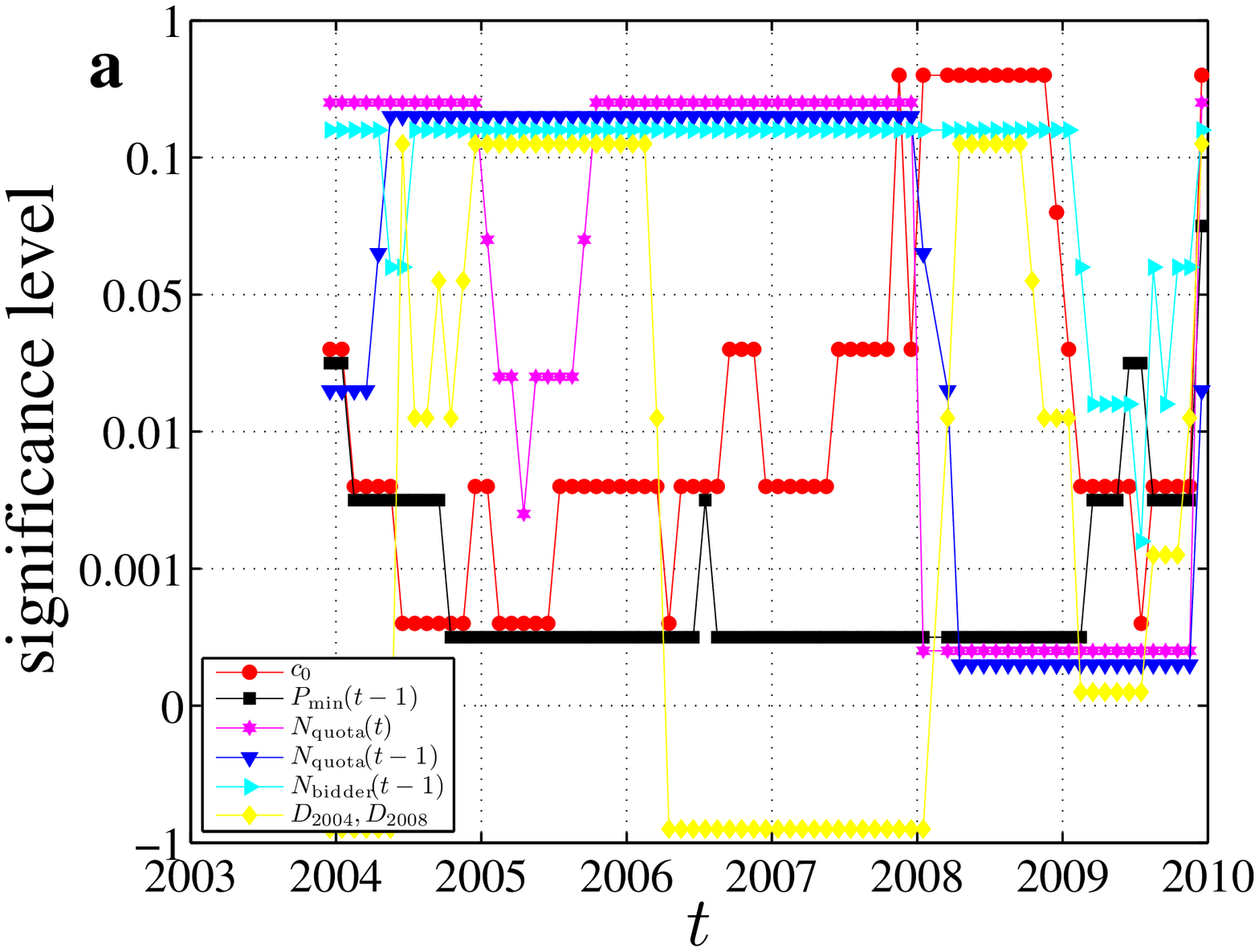}
 \includegraphics[width=8cm]{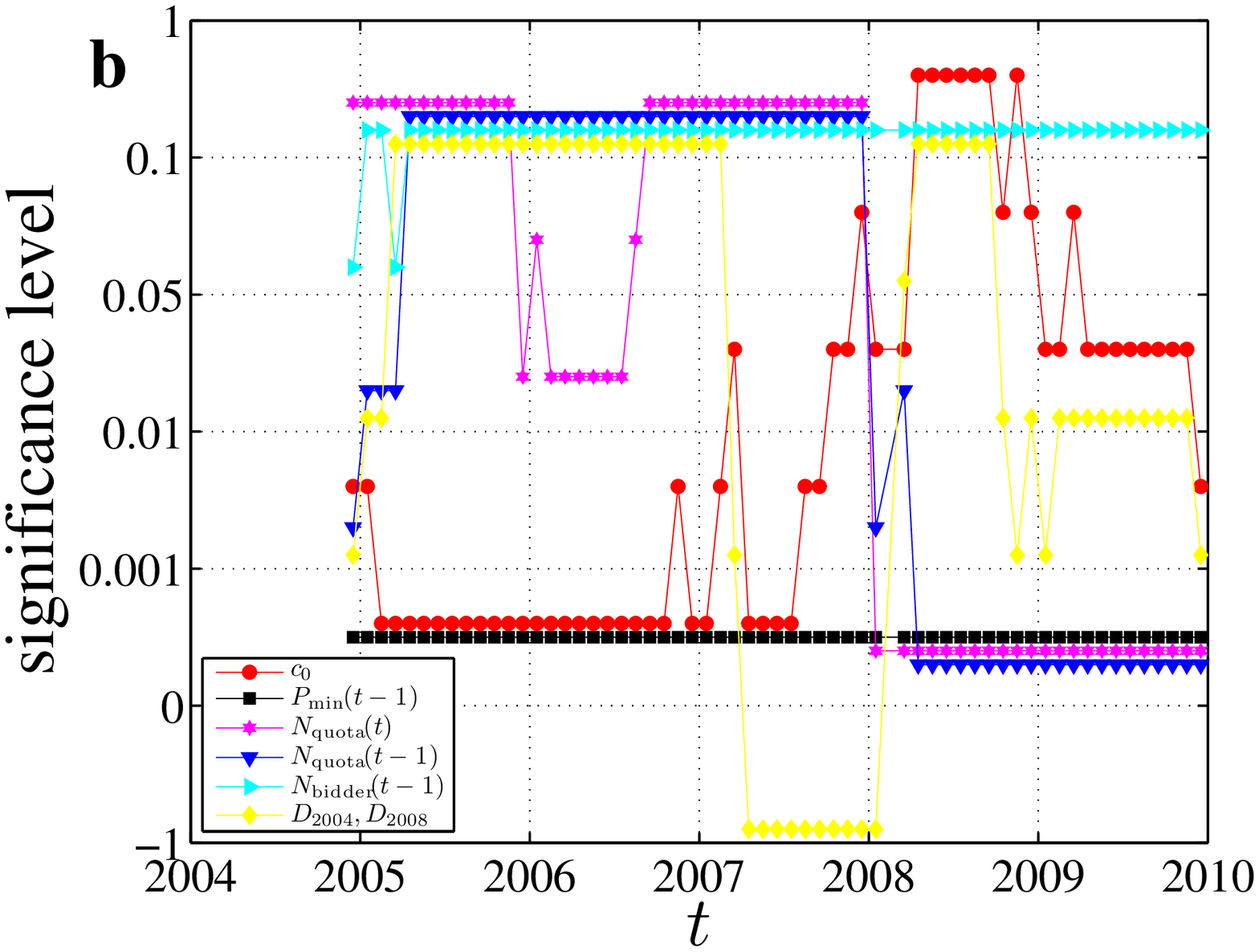}
 \caption{\label{Fig:SigLevel} (Color online) Evolving significance level of the influencing factors (or variables) for the moving window size $S=24$ (a) and $S=36$ (b). There are five intervals of significant level: $[0,0.001)$, $[0.001,0.01)$, $[0.01, 0.05)$, $[0.05,0.1)$, and $[0.1,1)$. The different locations of two points in a given significance interval are not related to the difference in their $p$-values. The bottom interval [-1,0) is not for significance level but to indicate that there is no dummy variable in the model for the associated window.}
\end{figure}

For each month $t$, we calibrate model (\ref{Eq:Model:3}) and obtain the $p$-values of the variables. For clarity, we use a quantitative presentation of the results in Fig.~\ref{Fig:SigLevel}. We divide the interval [0,1) into five subintervals $[s_1,s_2)$: $[0,0.001)$, $[0.001,0.01)$, $[0.01, 0.05)$, $[0.05,0.1)$, and $[0.1,1)$. Consider that the $p$-value of a given variable is $p$ at time $t$. If $s_1\leqslant{p}<s_2$, we plot a symbol in the interval $[s_1,s_2)$ in Fig.~\ref{Fig:SigLevel}. For the dummy variable, there are windows that do not contain either May 2004 or January 2008. In this case, neither of the two dummy variables are included in the regression and we plot points in the bottom interval $[-1,0)$ in Fig.~\ref{Fig:SigLevel}. It is necessary to point out that one dummy variable is included in the model only when May 2004 or January 2008 is not the at the edge of the window. Otherwise, the model is singular.

In general, the two plots in Fig.~\ref{Fig:SigLevel} are delivering very similar information. The results are summarized below for each variable.
\begin{itemize}
\item $P_{\min}(t-1)$: When $S=24$, the variable is significant at the 0.05 level for almost all months. In addition, the variable is significant at the 0.001 level for most months in years from 2005 to 2008 and significant at the 0.01 level for most months in years 2004 and 2009. When $S=36$, the variable is significant at the 0.001 level for all months under investigation.
\item $N_{\rm{quota}}(t)$: Basically, the variable is significant at the 0.001 level after January 2008 and insignificant since the $p$-values are greater than 0.1 before January 2008, which holds for both $S=24$ and $S=36$. For some of the months where May 2004 is within the window, the $p$-values are less than 0.05.
\item $N_{\rm{quota}}(t-1)$: Similar to $N_{\rm{quota}}(t)$, the variable is significance at the 0.001 level after January 2008 and insignificant with the $p$-values greater than 0.1 before January 2008, which holds for both $S=24$ and $S=36$ with a few exceptional months.
\item $N_{\rm{bidder}}(t-1)$: When $S=36$, the variable is insignificant at the 0.1 level for all months with two exceptions. When $S=24$, the variable is insignificant at the 0.1 level for all months in years from 2004 to 2008 with two exceptions. In 2009, we observe weak influence of this variable on the average price $P_{\rm{mean}}(t)$ and the variable is significant at the 0.05 level for six months.
\item $D_{2004}$: For this variable, we should focus on the ``diamonds'' in Fig.~\ref{Fig:SigLevel} before 2007. For both $S=24$ and $S=36$, the variable is insignificant at the 0.1 level for all months with only a few exceptions.
\item $D_{2008}$: For this variable, we should focus on the ``diamonds'' in Fig.~\ref{Fig:SigLevel} after 2007. Roughly speaking, for both $S=24$ and $S=36$, the variable is insignificant at the 0.1 level for the early months in 2008 and becomes significant at the 0.05 level afterwards. It shows that the change of the auction rules does have influence on the average price, which can be detected only when enough data points are included in the model.
\end{itemize}

According to these results, we should use different models for the data before and after January 2008. Before the change of auction rules, the model should contain only $P_{\min}(t-1)$ and $D_{2004}$. We can use the following
\begin{equation}
 P_{\rm{mean}}(t) = c_0+c_1P_{\min}(t-1) +d_1 D_{2004}+\epsilon(t),
 \label{Eq:Model:5}
\end{equation}
which is the model adopted in Ref.~\cite{Xu-2005-ET}. Since the influence of the dummy variable $D_{2004}$ is marginal, we should also consider the following model for comparison:
\begin{equation}
 P_{\rm{mean}}(t) = c_0+c_1P_{\min}(t-1)+\epsilon(t).
 \label{Eq:Model:6}
\end{equation}
Since January 2008, we should use model (\ref{Eq:Model:4}), as well as model (\ref{Eq:Model:2}) for comparison.

\begin{figure}[htb]
 \centering
 \includegraphics[width=8cm]{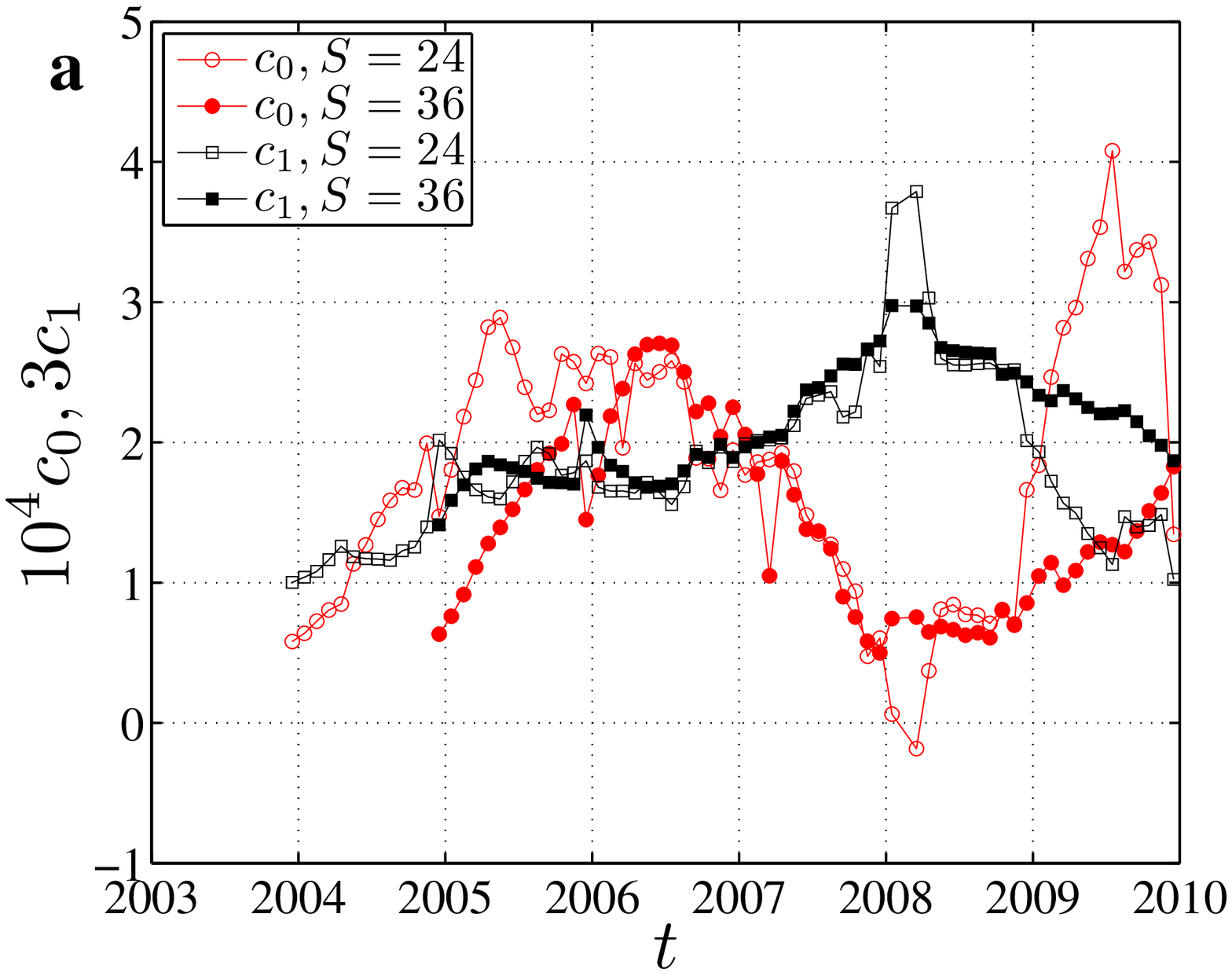}
 \includegraphics[width=8cm]{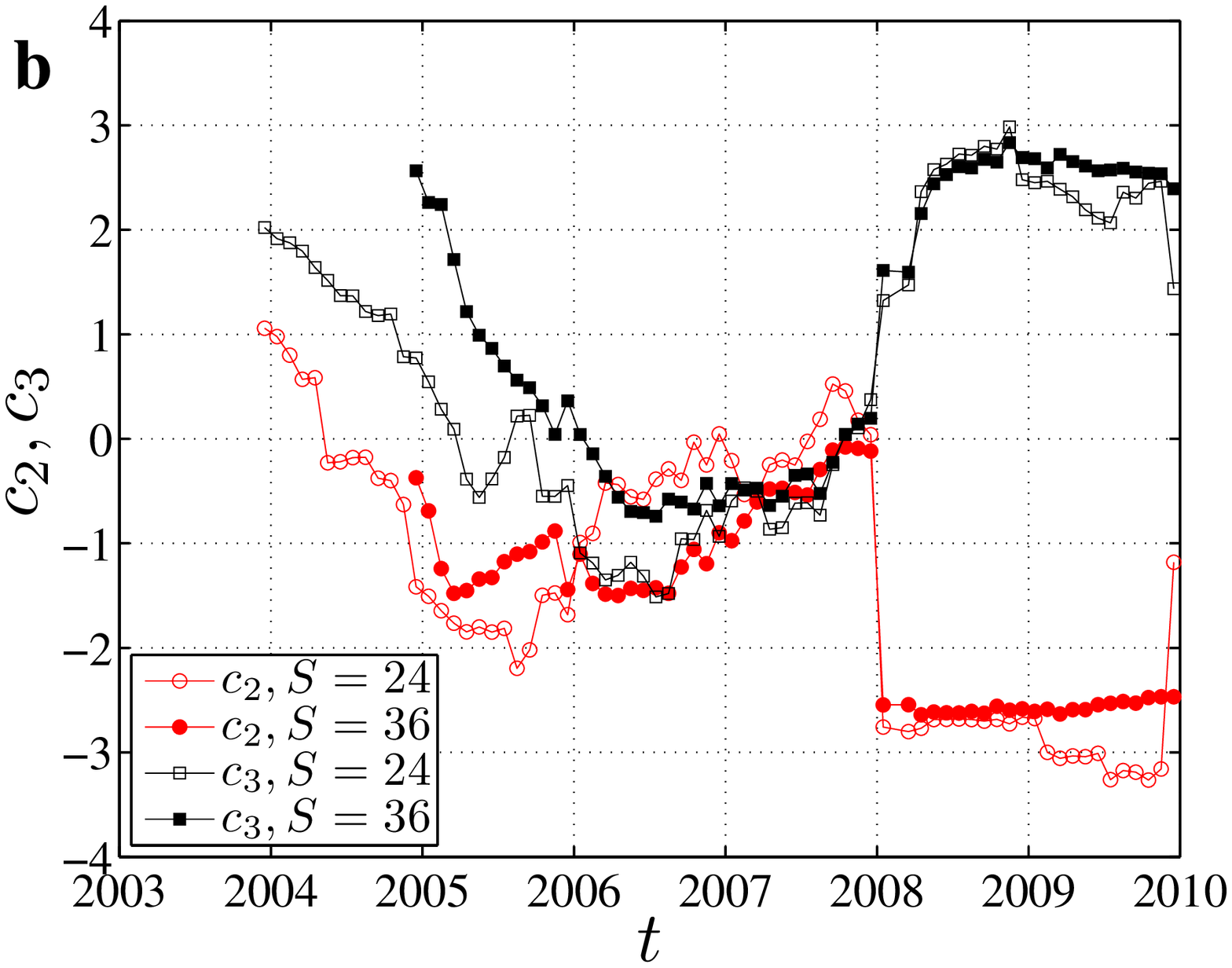}
 \caption{\label{Fig:c0c1c2c3} (Color online) Evolution of the estimated coefficients: (a) $c_0$ and $c_1$, (b) $c_2$ and $c_3$. For better presentation, we plot $10^4c_0$ and $2c_1$ instead of $c_0$ and $c_1$.}
\end{figure}

Figure \ref{Fig:c0c1c2c3} shows the evolution of the estimated coefficients $c_0$, $c_1$, $c_2$ and $c_3$. It is found that the two evolutionary curves for each coefficient are qualitatively the same. For all coefficients, we observe a change of regime around January 2008. As shown in Fig.~\ref{Fig:c0c1c2c3}a, the coefficient $c_1$ for the variable $P_{\min}$ increases before January 2008 and decreases afterwards. In Fig.~\ref{Fig:c0c1c2c3}b, we find that $c_2$ and $c_3$ evolved in a very similar way before January 2008 and then experienced a sudden change with a decrease in $c_2$ and an increase in $c_3$, which makes Fig.~\ref{Fig:c0c1c2c3}b look like a bifurcation diagram. Therefore, the average price $P_{\rm{mean}}(t)$ is negatively related to the quota $N_{\rm{quota}}(t)$ in the same month as expected and negatively related to the quota $N_{\rm{quota}}(t-1)$ in the preceding month. In addition, we find that $c_2\approx c_3$ so that model (\ref{Eq:Model:4}) can be rewritten as
\begin{equation}
 P_{\rm{mean}}(t) = c_0+c_1P_{\min}(t-1) - c_3\Delta{N_{\rm{quota}}} +d_2 D_{2008}+\epsilon(t),
 \label{Eq:Model:10}
\end{equation}
where $c_3>0$ and $\Delta{N_{\rm{quota}}} = N_{\rm{quota}}(t)-N_{\rm{quota}}(t-1)$ is the increment of quota compared with the preceding month. This model (\ref{Eq:Model:10}) has a more clear economic interpretation than model (\ref{Eq:Model:4}) that the average price is expect to decrease if more PCLP quota is released compared to that in the preceding month.

\subsection{Prediction}
\label{S2:Prediction}

\begin{figure}[htb]
 \centering
 \includegraphics[width=8cm]{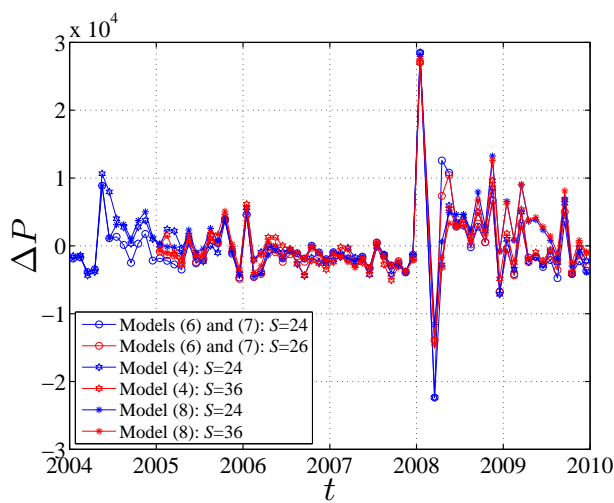}
 \caption{\label{Fig:Predict} (Color online) Evolution of the difference between the predicted price with reference to the true average price: $\Delta P(t) = \hat{P}_{\rm{mean}}(t) - P_{\rm{mean}}(t)$ for the three models: a mixed model of Eq.~(\ref{Eq:Model:4}) and Eq.~(\ref{Eq:Model:5}) with dummy variables, model (\ref{Eq:Model:2}), and model (\ref{Eq:Model:6}).}
\end{figure}

Based on the results of Section \ref{S2:Model4Predict}, we can make out-of-sample prediction of the average prices. For comparison, we use three predictive models: a mixed model of Eq.~(\ref{Eq:Model:4}) and Eq.~(\ref{Eq:Model:5}) with dummy variables, model (\ref{Eq:Model:2}), and model (\ref{Eq:Model:6}). For each model, we obtain the estimates of the parameters and extrapolate to obtain the predicted average price $\hat{P}_{\rm{mean}}(t+1)$ in the successive month. This is done for $S=24$ and $S=36$. Fig.~\ref{Fig:Predict} illustrates the difference between the predicted price with reference to the true average price:
\begin{equation}
 \Delta P(t) = \hat{P}_{\rm{mean}}(t) - P_{\rm{mean}}(t).
 \label{Eq:dP}
\end{equation}
The most striking feature of Fig.~\ref{Fig:Predict} is that the predictions for the prices in January and March of 2008 completely fail. Note that there was no auction in February 2008. We also find that the curves are more noisy after January 2008 than before, which means that the auction became more efficient after it became more transparent. However, this claim calls for more data since we observe a decreasing trend in $\Delta P(t)$.

Table \ref{TB:Prediction} shows the averages of absolute prediction errors $|\Delta{P}|$ for the three models and for different samples. Several features can be extracted. First, the results are similar for $S=24$ and $S=36$. However, each value for $S=24$ is larger than the corresponding value for $S=36$ implying that the prediction using a longer window $S=36$ systematically outperforms that with a shorter window $S=24$. Second, the values in column $B_1$ are significantly less than those in column $B_2$, which is a quantitative description that the curves in Fig.~\ref{Fig:c0c1c2c3} after January 2008 are more noisy. Third, for sample $B_1$, we find that model (\ref{Eq:Model:2}) has the worst predictive power for both $S=24$ and $S=36$, the mixed model (or equivalently model (\ref{Eq:Model:5})) outperforms model (\ref{Eq:Model:6}) when $S=24$, and model (\ref{Eq:Model:6}) outperforms the mixed model (or equivalently model (\ref{Eq:Model:5})) when $S=36$. Fourth, for sample $B_2$, we find that model (\ref{Eq:Model:2}) has the best predictive performance for both $S=24$ and $S=36$. The introducing of the dummy variable $D_{2008}$ deteriorates the predictions. In summary, we can conclude that the simpler models (\ref{Eq:Model:2}) and (\ref{Eq:Model:6}) outperform models (\ref{Eq:Model:4}) and (\ref{Eq:Model:5}) for samples $B_2$ and $B_1$, respectively. This is reminiscent of the conventional wisdom that the simpler is the better, which seems true in prediction \cite{Sornette-2003}.

\begin{table}[htb]
\centering
\caption{Average of the absolute prediction errors $|\Delta{P}|$ for the three models and for different samples. Sample $A$ contains all months from January 2004 to December 2009 for $S=24$ and from January 2005 to December 2009 for $S=36$. Sample $B$ contains all months in $A$ but discarding January 2008 and March 2008. Sample $B_1$ contains the months in $B$ before January 2008 and sample $B_2$ contains the months in $B$ after January 2008 so that $B=B_1\cup B_2$.}
\label{TB:Prediction}
\medskip
\begin{tabular}{ccccccccccccccccccccccccc}
  \hline\hline
   \multirow{3}*[2mm]{Model} & \multicolumn{4}{c}{$S=24$}&& \multicolumn{4}{c}{$S=36$} \\
  \cline{2-5}\cline{7-10}
                                           & $A$      & $B$      & $B_1$    & $B_2$   & & $A$      & $B$      & $B_1$    & $B_2$   \\\hline
 (\ref{Eq:Model:4}) and (\ref{Eq:Model:5}) & $ 3284$  & $ 2643$  & $\bf{2118}$  & $ 3842$ & & $ 3234$  & $ 2629$  & $ 2057$  & $ 3608$ \\
                        (\ref{Eq:Model:2}) & $ 3361$  & $ 2722$  & $ 2377$  & $\bf{3510}$ & & $ 3153$  & $ 2545$  & $ 2126$  & $\bf{3264}$ \\
                        (\ref{Eq:Model:6}) & $ 3230$  & $ 2750$  & $ 2170$  & $ 4076$ & & $ 3158$  & $ 2528$  & $\bf{1906}$  & $ 3596$ \\
  \hline\hline
\end{tabular}
\end{table}

\section{Summary}
\label{S1:Summay}

We have conducted statistical analysis of the average prices $P_{\rm{mean}}$ of the license plates of private cars in Shanghai. We consider three endogenous variables including the minimal price $P_{\min}$ of the bidding winners, the quota $N_{\rm{quota}}$ of private car license plates, and the number $N_{\rm{bidder}}$ of bidders, as well as two exogenous shocks known as the legality debate of the auction in May 2004 and the auction regime reform in January 2008. The data sets cover the period from January 2002 to December 2009. We found that the legality debate of the auction had marginal transient impact on the average price in a short time period, while the change of the auction rules has significant influence on the average price, which reduces the average price by about 3020 RMB. Therefore, the evolution of the average price entered a new regime caused by the auction reform in January 2008.

The number $N_{\rm{bidder}}$ of bidders is found to have no influence on the average price in the whole time period. Before January 2008, the average price $P_{\rm{mean}}$ was only influenced by the minimal price $P_{\min}$ in the preceding month with a positive correlation. The quotas in the nearest two months become additional significant influencing factors since January 2008. It means that the government is able to manipulate the average price by controlling the quota of license plate after the auction reform.

We compared the predictive power of several models using 2-year and 3-year moving windows. It is found that the two models without dummy variables had better performance and the use of the 3-year moving window gave better predictions on average. In addition, the average absolute prediction error in the second regime is about 1,358 RMB higher than in the first regime. We figure that the auction market becomes more efficient thanks to the auction reform in January 2008.

\bigskip
{\textbf{Acknowledgments:}}

This work was partially supported by the Program for New Century Excellent Talents in University (NCET-07-0288).

\bibliography{E:/Papers/Auxiliary/Bibliography}

\end{document}